# Randomization Does Not Justify Logistic Regression

**David A. Freedman**

*Abstract.* The logit model is often used to analyze experimental data. However, randomization does not justify the model, so the usual estimators can be inconsistent. A consistent estimator is proposed. Neyman's non-parametric setup is used as a benchmark. In this setup, each subject has two potential responses, one if treated and the other if untreated; only one of the two responses can be observed. Beside the mathematics, there are simulation results, a brief review of the literature, and some recommendations for practice.

*Key words and phrases:* Models, randomization, logistic regression, logit, average predicted probability.

## 1. INTRODUCTION

The logit model is often fitted to experimental data. As explained below, randomization does not justify the assumptions behind the model. Thus, the conventional estimator of log odds is difficult to interpret; an alternative will be suggested. Neyman's setup is used to define parameters and prove results. (Grammatical niceties apart, the terms "logit model" and "logistic regression" are used interchangeably.)

After explaining the models and estimators, we present simulations to illustrate the findings. A brief review of the literature describes the history and current usage. Some practical recommendations are derived from the theory. Analytic proofs are sketched at the end of the paper.

## 2. NEYMAN

There is a study population with $n$ subjects indexed by $i = 1, \ldots, n$. Fix $\pi_T$ with $0 < \pi_T < 1$. Choose

*David A. Freedman is Professor, Department of Statistics, University of California, Berkeley, California 94720-3860, USA e-mail: census@stat.berkeley.edu.*

$n\pi_T$ subjects at random and assign them to the treatment condition. The remaining $n\pi_C$ subjects are assigned to a control condition, where $\pi_C = 1 - \pi_T$. According to Neyman (1923), each subject has two responses: $Y_i^T$ if assigned to treatment, and $Y_i^C$ if assigned to control. The responses are 1 or 0, where 1 is "success" and 0 is "failure." Responses are fixed, that is, not random.

If $i$ is assigned to treatment $(T)$, then $Y_i^T$ is observed. Conversely, if $i$ is assigned to control $(C)$, then $Y_i^C$ is observed. Either one of the responses may be observed, but not both. Thus, responses are subject-level parameters. Even so, responses are estimable (see Section 9). Each subject has a covariate $Z_i$, unaffected by assignment; $Z_i$ is observable. In this setup, the only stochastic element is the randomization: conditional on the assignment variable $X_i$, the observed response $Y_i = X_i Y_i^T + (1 - X_i) Y_i^C$ is deterministic.

Population-level ITT (intention-to-treat) parameters are defined by taking averages over all $n$ subjects in the study population:

$$
\begin{aligned}
\alpha^T &= \frac{1}{n} \sum Y_i^T, \\
\alpha^C &= \frac{1}{n} \sum Y_i^C.
\end{aligned}
\quad (1)
$$

For example, $\alpha^T$ is the fraction of successes if all subjects are assigned to $T$; similarly for $\alpha^C$. A parameter of considerable interest is the differential log odds of success,

$$(2) \qquad \Delta = \log \frac{\alpha^T}{1-\alpha^T} - \log \frac{\alpha^C}{1-\alpha^C}.$$

The logit model is all about log odds (more on this below). The parameter $\Delta$ defined by (2) may therefore be what investigators think is estimated by running logistic regressions on experimental data, although that idea is seldom explicit.

### The Intention-to-Treat Principle

The intention-to-treat principle, which goes back to Hill (1961, page 259), is to make comparisons based on treatment assigned rather than treatment received. Such comparisons take full advantage of the randomization, thereby avoiding biases due to self-selection. For example, the unbiased estimators for the parameters in (1) are the fraction of successes in the treatment group and the control group, respectively. Below, these will be called *ITT estimators*. ITT estimators measure the effect of assignment rather than treatment. With crossover, the distinction matters. For additional discussion, see Freedman (2006a).

## 3. THE LOGIT MODEL

To set up the logit model, we consider a study population of $n$ subjects, indexed by $i = 1, \ldots, n$. Each subject has three observable random variables: $Y_i, X_i, Z_i$. Here, $Y_i$ is the response, which is 0 or 1. The primary interest is the "effect" of $X_i$ on $Y_i$, and $Z_i$ is a covariate.

For our purposes, the best way to formulate the model involves a latent (unobservable) random variable $U_i$ for each subject. These are assumed to be independent across subjects, with a common *logistic distribution*: for $-\infty < u < \infty$,

$$(3) \qquad P(U_i < u) = \exp(u)/[1 + \exp(u)],$$

where $\exp(u) = e^u$. The model assumes that $X$ and $Z$ are *exogenous*, that is, independent of $U$. More formally, $\{X_i, Z_i : i = 1, \ldots, n\}$ is assumed to independent of $\{U_i : i = 1, \ldots, n\}$. Finally, the model assumes that $Y_i = 1$ if

$$\beta_1 + \beta_2 X_i + \beta_3 Z_i + U_i > 0;$$

else, $Y_i = 0$.

Given $X$ and $Z$, it follows that responses are independent across subjects, the conditional probability that $Y_i = 1$ being $p(\beta, X_i, Z_i)$, where

$$(4) \qquad p(\beta, x, z) = \frac{\exp(\beta_1 + \beta_2 x + \beta_3 z)}{1 + \exp(\beta_1 + \beta_2 x + \beta_3 z)}.$$

(To verify this, check first that $-U_i$ is distributed like $+U_i$.) The parameter vector $\beta = (\beta_1, \beta_2, \beta_3)$ is usually estimated by maximum likelihood. We denote the MLE by $\hat{\beta}$.

### Interpreting the Coefficients in the Model

In the case of primary interest, $X_i$ is 1 or 0. Consider the log odds $\lambda_i^T$ of success when $X_i = 1$, as well as the log odds $\lambda_i^C$ when $X_i = 0$. In view of (4),

$$
\begin{aligned}
\lambda_i^T &= \log \frac{p(\beta, 1, Z_i)}{1 - p(\beta, 1, Z_i)} \\
&= \beta_1 + \beta_2 + \beta_3 Z_i, \\
\lambda_i^C &= \log \frac{p(\beta, 0, Z_i)}{1 - p(\beta, 0, Z_i)} \\
&= \beta_1 + \beta_3 Z_i.
\end{aligned}
\tag{5}
$$

In particular, $\lambda_i^T - \lambda_i^C = \beta_2$ for all $i$, whatever the value of $Z_i$ may be. Thus, according to the model, $X_i = 1$ adds $\beta_2$ to the log odds of success.

### Application to Experimental Data

To apply the model to experimental data, define $X_i = 1$ if $i$ is assigned to $T$, while $X_i = 0$ if $i$ assigned to $C$. Notice that the model not justified by randomization. Why would the logit specification be correct rather than the probit—or anything else? What justifies the choice of covariates? Why are they exogenous? If the model is wrong, what is $\hat{\beta}_2$ supposed to be estimating? The last rhetorical question may have an answer: the parameter $\Delta$ in (2) seems like a natural choice, as indicated above.

More technically, from Neyman's perspective, given the assignment variables $\{X_i\}$, the responses are deterministic: $Y_i = Y_i^T$ if $X_i = 1$, while $Y_i = Y_i^C$ if $X_i = 0$. The logit model, on the other hand, views the responses $\{Y_i\}$ as random—with a specified distribution—given the assignment variables and covariates.

The contrast is therefore between two styles of inference.

- Randomization provides a known distribution for the assignment variables; statistical inferences are based on this distribution.
- Modeling assumes a distribution for the latent variables; statistical inferences are based on that assumption. Furthermore, model-based inferences are conditional on the assignment variables and covariates.

A similar contrast will be found in other areas too, including sample surveys. See Koch and Gillings (2005) for a review and pointers to the literature.

### What if the Logit Model is Right?

Suppose the model is right, and there is a causal interpretation. We can intervene and set $X_i$ to 1 without changing the $Z$'s or $U$'s, so $Y_i = 1$ if and only if $\beta_1 + \beta_2 + \beta_3 Z_i + U_i > 0$. Similarly, we can set $X_i$ to 0 without changing anything else, and then $Y_i = 1$ if and only if $\beta_1 + \beta_3 Z_i + U_i > 0$. Notice that $\beta_2$ appears when $X_i$ is set to 1, but disappears when $X_i$ is set to 0.

On this basis, for each subject, whatever the value of $Z_i$ may be, setting $X_i$ to 1 rather than 0 adds $\beta_2$ to the log odds of success. If the model is right, $\beta_2$ is a very useful parameter, which is well estimated by the MLE provided $n$ is large. For additional detail on causal modeling and estimation, see Freedman (2005).

Even if the model is right and $n$ is large, $\beta_2$ differs from $\Delta$ in (2). For instance, $\alpha^T$ will be nearly equal to $\frac{1}{n}\sum_{i=1}^{n} p(\beta, 1, Z_i)$. So $\log \alpha^T - \log(1 - \alpha^T)$ will be nearly equal to

$$
(6) \quad \log\left(\frac{1}{n}\sum_{i=1}^{n} p(\beta, 1, Z_i)\right) - \log\left(\frac{1}{n}\sum_{i=1}^{n}[1 - p(\beta, 1, Z_i)]\right).
$$

Likewise, $\log \alpha^C - \log(1 - \alpha^C)$ will be nearly equal to

$$
(7) \quad \log\left(\frac{1}{n}\sum_{i=1}^{n} p(\beta, 0, Z_i)\right) - \log\left(\frac{1}{n}\sum_{i=1}^{n}[1 - p(\beta, 0, Z_i)]\right).
$$

Taking the log of an average, however, is quite different from taking the average of the logs. The former is relevant for $\Delta$ in (2), as shown by (6)–(7); the latter for computing

$$
(8) \quad \frac{1}{n}\sum_{i=1}^{n}(\lambda_i^T - \lambda_i^C) = \beta_2,
$$

where the log odds of success $\lambda_i^T$ and $\lambda_i^C$ were computed in (5).

The difference between averaging inside and outside the logs may be surprising at first, but in the end, that difference is why you should put confounders like $Z$ into the equation—if you believe the model. Section 9 below gives further detail, and an inequality relating $\beta_2$ to $\Delta$.

### From Neyman to Logits

How could we get from Neyman to the logit model? To begin with, we would allow $Y_i^T$ and $Y_i^C$ to be 0–1 valued random variables; the $Z_i$ can be random too. To define the parameters in (1) and (2), we would replace $Y_i^T$ and $Y_i^C$ by their expectations. None of this is problematic, and the Neyman model is now extremely general and flexible. Randomization makes the assignment variables $\{X_i\}$ independent of the potential responses $Y_i^T, Y_i^C$.

To get the logit model, however, we would need to specialize this setup considerably, assuming the existence of IID logistic random variables $U_i$, independent of the covariates $Z_i$, with

$$
(9) \quad \begin{aligned} Y_i^T &= 1 \quad \text{if and only if} \\ &\quad \beta_1 + \beta_2 + \beta_3 Z_i + U_i > 0, \\ Y_i^C &= 1 \quad \text{if and only if} \\ &\quad \beta_1 + \beta_3 Z_i + U_i > 0. \end{aligned}
$$

Besides (9), the restrictive assumptions are the following:

 (i) The $U_i$ are independent of the $Z_i$.
 (ii) The $U_i$ are independent across subjects $i$.
 (iii) The $U_i$ have a common logistic distribution.

If you are willing to make these assumptions, what randomization contributes is a guarantee that the assignment variables $\{X_i\}$ are independent of the latent variables $\{U_i\}$. Randomization does not guarantee the existence of the $U_i$, or the truth of (9), or the validity of (i)–(iii).

## 4. A PLUG-IN ESTIMATOR FOR THE LOG ODDS

If a logit model is fitted to experimental data, average predicted probabilities are computed by plugging $\hat{\beta}$ into (4):

$$
(10a) \quad \begin{aligned} \tilde{\alpha}^T &= \frac{1}{n}\sum_{i=1}^{n} p(\hat{\beta}, 1, Z_i), \\ \tilde{\alpha}^C &= \frac{1}{n}\sum_{i=1}^{n} p(\hat{\beta}, 0, Z_i). \end{aligned}
$$

(The tilde notation is needed; $\hat{\alpha}^T$ and $\hat{\alpha}^C$ will make their appearances momentarily.) Then the differential log odds in (2) can be estimated by plugging into the formula for $\Delta$:

$$
(10b) \quad \tilde{\Delta} = \log\frac{\tilde{\alpha}^T}{1-\tilde{\alpha}^T} - \log\frac{\tilde{\alpha}^C}{1-\tilde{\alpha}^C}.
$$

As will be seen below, $\tilde{\Delta}$ is consistent.

The ITT (intention-to-treat) estimators are defined as follows:

$$\text{(11a)} \quad \hat{\alpha}^T = \frac{1}{n_T} \sum_{i \in T} Y_i, \quad \hat{\alpha}^C = \frac{1}{n_C} \sum_{i \in C} Y_i,$$

where $n_T = n\pi_T$ is the number of subjects in $T$ and $n_C = n\pi_C$ is the number of subjects in $C$. Then

$$\text{(11b)} \quad \hat{\Delta} = \log \frac{\hat{\alpha}^T}{1 - \hat{\alpha}^T} - \log \frac{\hat{\alpha}^C}{1 - \hat{\alpha}^C}.$$

The ITT estimators are consistent too, with asymptotics discussed in Freedman (2008a, 2008b). The intuition: $\hat{\alpha}^T$ is the average success rate in the treatment group, and the sample average is a good estimator for the population average. The same reasoning applies to $\hat{\alpha}^C$.

## 5. SIMULATIONS

The simulations in this section are designed to show what happens when the logit model is fitted to experimental data. The data generating mechanism is not the logit, so the simulations illustrate the consequences of specification error. The stochastic element is the randomization, as in Section 2. (Some auxiliary randomness is introduced to construct the individual-level parameters, but that gets conditioned away.) Let $n = 100, 500, 1000, 5000$. For $i = 1, \ldots, n$:
  let $U_i, V_i$ be IID uniform random variables,
  let $Z_i = V_i$,
  let $Y_i^C = 1$ if $U_i > 1/2$, else $Y_i^C = 0$,
  let $Y_i^T = 1$ if $U_i + V_i > 3/4$, else $Y_i^T = 0$.

Suppose $n$ is very large. The mean response in the control condition is around $P(U_i > 1/2) = 1/2$, so the odds of success in the control condition are around 1. (The qualifiers are needed because the $U_i$ are chosen at random.) The mean response in the treatment condition is around 23/32, because

$$P(U_i + V_i < 3/4) = (1/2) \times (3/4)^2 = 9/32.$$

So the odds of success in the treatment condition are around $(23/32)/(9/32)$. The parameter $\Delta$ in (2) will therefore be around

$$\log \frac{23/32}{9/32} - \log 1 = \log \frac{23}{9} = 0.938.$$

Even for moderately large $n$, non-linearity in (2) is an issue, and the approximation given for $\Delta$ is unsatisfactory.

TABLE 1
*Simulations for $n = 100$, 500, 1000, 5000. Twenty-five percent of the subjects are assigned at random to $C$, the rest to $T$. Averages and SDs are shown for the MLE $\hat{\beta}$ and the plug-in estimator $\tilde{\Delta}$, as well as the true value of the differential log odds $\Delta$ defined in (2). There are 1,000 simulated experiments for each $n$*

| $n$ | $\hat{\beta}_1$ | $\hat{\beta}_2$ | $\hat{\beta}_3$ | Plug-in | Truth |
|---|---|---|---|---|---|
| 100 | −0.699 | 1.344 | 2.327 | 1.248 | 1.245 |
|     | 0.457  | 0.540 | 0.621 | 0.499 |       |
| 500 | −1.750 | 1.263 | 3.318 | 1.053 | 1.053 |
|     | 0.214  | 0.234 | 0.227 | 0.194 |       |
| 1000| −1.568 | 1.046 | 3.173 | 0.885 | 0.883 |
|     | 0.155  | 0.169 | 0.154 | 0.142 |       |
| 5000| −1.676 | 1.134 | 3.333 | 0.937 | 0.939 |
|     | 0.071  | 0.076 | 0.072 | 0.062 |       |

The construction produces individual-level variation: a majority of subjects are unaffected by treatment, about 1/4 are helped, about 1/32 are harmed. The covariate is reasonably informative about the effect of treatment—if $Z_i$ is big, treatment is likely to help.

Having constructed $Z_i$, $Y_i^C$ and $Y_i^T$ for $i = 1, \ldots, n$, we freeze them, and simulate 1000 randomized controlled experiments, where 25% of the subjects are assigned to $C$ and 75% to $T$. We fit a logit model to the data generated by each experiment, computing the MLE $\hat{\beta}$ and the plug-in estimator $\tilde{\Delta}$ defined by (10b). The average of the 1000 $\hat{\beta}$'s and $\tilde{\Delta}$'s is shown in Table 1, along with the true value of the differential log odds, namely, $\Delta$ in (2). We distinguish between the standard deviation (SD) and the standard error (SE). Below each average, the table shows the corresponding SD.

For example, with $n = 100$, the average of the 1000 $\hat{\beta}_2$'s is 1.344; the SD is 0.540; the Monte Carlo SE in the average is therefore $0.540/\sqrt{1000} = 0.017$. The average of the 1000 plug-in estimates is 1.248, and the true $\Delta$ is 1.245. When $n = 5000$, the bias in $\hat{\beta}_2$ as an estimator of $\Delta$ is $1.134 - 0.939 = 0.195$, with a Monte Carlo SE of $0.076/\sqrt{1000} = 0.002$. There is a confusion to avoid: $n$ is the number of subjects in the study population, varying from 100 to 5000, but the number of simulated experiments is fixed at 1000. (The *Monte Carlo SE* measures the impact of randomness in the simulation, which is based on a sample of "only" 1000 observations.)

The plug-in estimator is essentially unbiased and less variable than $\hat{\beta}_2$. The true value of $\Delta$ changes from one $n$ to the next, since values of $Y_i^C, Y_i^T$ are generated by Monte Carlo for each $n$. Even with $n = 5000$, the true value of $\Delta$ would change from one run to another, the SD across runs being about 0.03 (not shown in the table).

Parameter choices—for instance, the joint distribution of $(U_i, V_i)$—were somewhat arbitrary. Surprisingly, bias depends on the fraction of subjects assigned to $T$. On the other hand, changing the cutpoints used to define $Y_i^C$ and $Y_i^T$ from 1/2 and 3/4 to 0.95 and 1.5 makes little difference to the performance of $\hat{\beta}_2$ and the plug-in estimator. In these examples, the plug-in estimator and the ITT estimators are essentially unbiased; the latter has slightly smaller variance.

The bias in $\hat{\beta}_2$ depends very much on the covariate. For instance, if the covariate is $U_i + V_i$ rather than $V_i$, then $\hat{\beta}_2$ hovers around 3. Truth remains in the vicinity of 1, so the bias in $\hat{\beta}_2$ is huge. The plug-in and ITT estimators remain essentially unbiased, with variances much smaller than $\hat{\beta}_2$; the ITT estimator has higher variance than the plug-in estimator (data not shown for variations on the basic setup, or ITT estimators).

The Monte Carlo results suggest the following:

(i) As $n$ gets large, the MLE $\hat{\beta}$ stabilizes.
(ii) The plug-in estimator $\tilde{\Delta}$ is a good estimator of the differential log odds $\Delta$.
(iii) $\hat{\beta}_2$ tends to over-estimate $\Delta > 0$.

These points will be verified analytically below.

## 6. EXTENSIONS AND IMPLICATIONS

Suppose the differential log odds of success is the parameter to be estimated. Then $\hat{\beta}_2$ is generally the wrong estimator to use—whether the logit model is right or the logit model is wrong (Section 9 has a mathematical proof). It is better to use the plug-in estimator (10) or the ITT estimator (11). These estimators are nearly unbiased, and in many examples have smaller variances too.

Although details remain to be checked, the convergence arguments in Section 8 seem to extend to probits, the parameter corresponding to (2) being

$$\Phi^{-1}(\alpha^T) - \Phi^{-1}(\alpha^C),$$

where $\Phi$ is the standard normal distribution function. On the other hand, with the probit, the plug-in estimators are unlikely to be consistent, since the analogs of the likelihood equations (16)–(18) below involve weighted averages rather than simple averages.

In simulation studies (not reported here), the probit behaves very much like the logit, with the usual difference in scale: probit coefficients are about 5/8 of their logit counterparts (Amemiya, 1981, page 1487). Numerical calculations also confirm inconsistency of the plug-in estimators, although the asymptotic bias is small.

According to the logit and probit models, if treatment improves the chances of success, it does so for all subjects. In reality, of course, treatment may help some subgroups and hurt others. Subgroup analysis can therefore be a useful check on the models. Consistency of the plug-in estimators—as defined here—does not preclude subgroup effects.

Logit models, probit models, and their ilk are not justified by randomization. This has implications for practice. Rates and averages for the treatment and control groups should be compared before the modeling starts. If the models change the substantive results, that raises questions that need to be addressed.

There may be an objection that models take advantage of additional information. The objection has some merit *if* the models are right or nearly right. On the other hand, if the models cannot be validated, conclusions drawn from them must be shaky. "Cross-tabulation before regression" is a slogan to be considered.

## 7. LITERATURE REVIEW

Logit and probit models are often used to analyze experimental data. See Pate and Hamilton (1992), Gilens (2001), Hu (2003), Duch and Palmer (2004), Frey and Meier (2004), Gertler (2004). The plug-in estimator discussed here is similar to the "average treatment effect" sometimes reported in the literature; see, for example, Evans and Schwab (1995). For additional discussion, see Lane and Nelder (1982), Brant (1996).

Lim (1999) conjectured that plug-in estimators based on the logit model would be consistent, with an informal argument based on the likelihood equation. He also conjectured inconsistency for the probit. Middleton (2007) discusses inconsistent logit estimators.

The logistic distribution may first have been used to model population growth. See Verhulst (1845) and Yule (1925). Later, the distribution was used to model dose-response in bioassays (Berkson, 1944). An early biomedical application to causal inference is Truett, Cornfield, and Kannel (1967). The history is considered further in Freedman (2005). The present paper extends previous results on linear regression (Freedman, 2008a, 2008b).

Statistical models for causation go back to Jerzy Neyman's work on agricultural experiments in the early part of the 20th century. The key paper, Neyman (1923), was in Polish. There was an extended discussion by Scheffé (1956), and an English translation by Dabrowska and Speed (1990). The model was covered in elementary textbooks in the 1960s; see, for instance, Hodges and Lehmann (1964, Section 9.4). The setup is often called "Rubin's model," due in part to Holland (1986); that mistakes the history.

Neyman, Kolodziejczyk, and Iwaszkiewicz (1935) develop models with subject-specific random effects that depend on assignment, the objective being to estimate average expected values under various circumstances. This is discussed in Section 4 of Scheffé (1956).

Heckman (2000) explains the role of potential outcomes in econometrics. In epidemiology, a good source is Robins (1999). Rosenbaum (2002) proposes using models and permutation tests as devices for hypothesis-testing. This avoids difficulties outlined here: (i) if treatment has no effect, then $Y_i^T = Y_i^C = Y_i$ for all $i$, and (ii) randomization makes all permutations of $i$ equally likely—which is just what permutation tests need.

Rosenblum and van der Laan (2008) suggest that, at least for purposes of hypothesis testing, robust SEs will fix problems created by specification error. Such optimism is unwarranted. Under the alternative hypothesis, the robust SE is unsatisfactory because it ignores bias (Freedman, 2006b).

Under the null hypothesis, the robust SE may be asymptotically correct, but using it can reduce power (Freedman, 2008a, 2008b). In any event, if the null hypothesis is to be tested using model-based adjustments, exact $P$-values can be computed by permutation methods, as suggested by Rosenbaum (2002).

Models are often deployed to infer causation from association. For a discussion from various perspectives, see Berk (2004), Brady and Collier (2004), and Freedman (2005). The last summarizes a cross-section of the literature on this topic (pages 192–200).

Consider a logit model like the one in Section 3. Omitting the covariate $Z$ from the equation is called *marginalizing* over $Z$. The model is *collapsible* if the marginal model is again logit with the same $\beta_2$. In other words, given the $X$'s, the $Y$'s are conditionally independent, and

$$P(Y_i = 1|X_i) = \frac{\exp(\beta_1 + \beta_2 X_i)}{1 + \exp(\beta_1 + \beta_2 X_i)}.$$

Guo and Geng (1995) give conditions for collapsibility; also see Ducharme and Lepage (1986). Gail (1986, 1988) discusses collapsing when a design is balanced. Robinson and Jewell (1991) show that collapsing will usually decrease variance: logit models differ from linear models. Aris et al. (2000) review the literature and consider modeling strategies to compensate for non-collapsibility.

## 8. SKETCH OF PROOFS

We are fitting the logit model, which is incorrect, to data from an experiment. As before, let $X_i$ be the assignment variable, so $X_i = 1$ if $i \in T$ and $X_i = 0$ if $i \in C$. Let $Y_i$ be the observed response, so $Y_i = X_i Y_i^T + (1 - X_i) Y_i^C$. Let $L_n(\beta)$ be the "log-likelihood function" to be maximized. The quote marks are there because the model is wrong; $L_n$ is therefore only a pseudo-log-likelihood function. Abbreviate $p_i(\beta)$ for $p(\beta, X_i, Z_i)$ in (4). The formula for $L_n(\beta)$ is this:

(12a) $$L_n(\beta) = \sum_{i=1}^n T_i,$$

where

(12b) $$\begin{aligned} T_i &= \log[1 - p_i(\beta)] \\ &\quad + (\beta_1 + \beta_2 X_i + \beta_3 Z_i) Y_i. \end{aligned}$$

(The $T$ is for term, not treatment.) It takes a moment to verify (12), starting from the equation

(13) $$T_i = Y_i \log(p_i) + (1 - Y_i) \log(1 - p_i).$$

Each $T_i$ is negative. The function $\beta \to L_n(\beta)$ is strictly concave, as one sees by proving that $L_n''$ is a negative definite matrix. Consequently, there is a unique maximum at the MLE $\hat{\beta}_n$. We write $\hat{\beta}_n$ to show dependence on the size $n$ of the study population, although that creates a conflict in the notation. If pressed, we could write $\hat{\beta}_{n,j}$ for the $j$th component of the MLE.

The $i$th row of the "design matrix" is $(1, X_i, Z_i)$. Tacitly, we are assuming this matrix is nonsingular. For large $n$, the assumption will follow from regularity conditions to be imposed. The concavity of $L_n$ is well known. See, for instance, pages 122–123 in Freedman (2005) or page 273 in Amemiya (1985). Pratt (1981) discusses the history and proves a more general result.

For reference, we record one variation on these ideas. Let $M$ be an $n \times p$ matrix of rank $p$; write $M_i$ for the $i$th row of $M$. Let $y$ be an $n \times 1$ vector of 0s and 1s. Let $\beta$ be a $p \times 1$ vector. Let $w_i > 0$ for $i = 1, \ldots, n$. Consider $M$ and $y$ as fixed, $\beta$ as variable. Define $L(\beta)$ as

$$\sum_{i=1}^{n} w_i \{-\log[1 + \exp(M_i \cdot \beta)] + (M_i \cdot \beta) y_i\}.$$

PROPOSITION 1. *The function $\beta \to L(\beta)$ is strictly concave.*

One objective in the rest of this section is showing that

(14)  $\beta_n$ converges to a limit $\beta_\infty$ as $n \to \infty$.

A second objective is showing that

(15)  the plug-in estimator $\tilde{\Delta}$ is consistent.

The argument actually shows a little more. The plug-in estimator $\tilde{\alpha}^T$, the ITT estimator $\hat{\alpha}^T$, and the parameter $\alpha^T$ become indistinguishable as the size $n$ of the study population grows; likewise for $\tilde{\alpha}^C$, $\hat{\alpha}^C$ and $\alpha^C$.

The ITT estimators $\hat{\alpha}^T$, $\hat{\alpha}^C$ were defined in (11). Recall too that $n_T = n\pi_T$ and $n_C = n\pi_C$ are the numbers of subjects in $T$ and $C$ respectively. The statement of Lemma 1 involves the *empirical distribution* of $Z_i$ for $i \in T$, which assigns mass $1/n_T$ to $Z_i$ for each $i \in T$. Similarly, the empirical distribution of $Z_i$ for $i \in C$ assigns mass $1/n_C$ to $Z_i$ for each $i \in C$.

To prove Lemma 1, we need the likelihood equation $L'_n(\beta) = 0$. This vector equation unpacks to three scalar equations in three unknowns, the components of $\beta$ that make up $\hat{\beta}_n$:

(16)  $\quad \dfrac{1}{n_T} \sum_{i \in T} p(\hat{\beta}_n, 1, Z_i) = \dfrac{1}{n_T} \sum_{i \in T} Y_i,$

(17)  $\quad \dfrac{1}{n_C} \sum_{i \in C} p(\hat{\beta}_n, 0, Z_i) = \dfrac{1}{n_C} \sum_{i \in C} Y_i,$

(18)  $\quad \dfrac{1}{n} \sum_{i=1}^{n} p(\hat{\beta}_n, X_i, Z_i) Z_i = \dfrac{1}{n} \sum_{i=1}^{n} Y_i Z_i.$

This follows from (12)–(13) after differentiating with respect to $\beta_1$, $\beta_2$, and $\beta_3$—and then doing a bit of algebra.

LEMMA 1. *If the empirical distribution of $Z_i$ for $i \in T$ matches the empirical distribution for $i \in C$ (the first balance condition), then the plug-in estimators $\tilde{\alpha}^T$, $\tilde{\alpha}^C$ match the ITT estimators. More explicitly,*

$$\frac{1}{n} \sum_{i=1}^{n} p(\hat{\beta}_n, 1, Z_i) = \frac{1}{n_T} \sum_{i \in T} Y_i,$$

$$\frac{1}{n} \sum_{i=1}^{n} p(\hat{\beta}_n, 0, Z_i) = \frac{1}{n_C} \sum_{i \in C} Y_i.$$

PROOF. The plug-in estimators $\tilde{\alpha}^T$, $\tilde{\alpha}^C$ were defined in (10); the ITT estimators $\hat{\alpha}^T$, $\hat{\alpha}^C$, in (11). We begin with $\tilde{\alpha}^T$. By (16),

$$\frac{1}{n_T} \sum_{i \in T} p(\hat{\beta}_n, 1, Z_i) = \frac{1}{n_T} \sum_{i \in T} Y_i = \hat{\alpha}^T.$$

By the balance condition,

$$\frac{1}{n_C} \sum_{i \in C} p(\hat{\beta}_n, 1, Z_i) = \frac{1}{n_T} \sum_{i \in T} p(\hat{\beta}_n, 1, Z_i)$$

equals $\hat{\alpha}^T$ too. Finally, the average of $p(\hat{\beta}_n, 1, Z_i)$ over all $i$ is a mixture of the averages over $T$ and $C$. So $\tilde{\alpha}^T = \hat{\alpha}^T$ as required. The same argument works for $\tilde{\alpha}^C$, using (17). $\square$

For the next lemma, recall $\alpha^T$, $\alpha^C$ from (1). The easy proof is omitted, being very similar to the proof of the previous result.

LEMMA 2. *Suppose the empirical distribution of the pairs $(Y_i^T, Y_i^C)$ for $i \in T$ matches the empirical distribution for $i \in C$ (the second balance condition). Then $\hat{\alpha}^T = \alpha^T$ and $\hat{\alpha}^C = \alpha^C$.*

LEMMA 3. *Let $x$ be any real number. Then*

$$e^x - \tfrac{1}{2} e^{2x} < \log(1 + e^x) < e^x,$$

$$x + e^{-x} - \tfrac{1}{2} e^{-2x} < \log(1 + e^x) < x + e^{-x}.$$

The first bound is useful when $x$ is large and negative; the second, when $x$ is large and positive. To get the second bound from the first, write $1 + e^x = e^x(1 + e^{-x})$, then replace $x$ by $-x$. The first bound will look more familiar on substituting $y = e^x$. The proof is omitted, being "just" calculus.

For the next result, let $G$ be an open, bounded, convex subset of Euclidean space. Let $f_n$ be a strictly concave function on $G$, converging uniformly to $f_\infty$, which is also strictly concave. Let $f_n$ take its maximum at $x_n$, while $f_\infty$ takes its maximum at $x_\infty \in G$. Although the lemma is well known, a proof may be helpful. We write $G \setminus H$ for the set of points that are in $G$ but not in $H$.

LEMMA 4. $x_n \to x_\infty$ and $f_n(x_n) \to f_\infty(x_\infty)$.

PROOF. Choose a small neighborhood $H$ of $x_\infty = \arg\max f_\infty$. There is a small positive $\delta$ with $f_\infty(x) < f_\infty(x_\infty) - \delta$ for $x \in G \setminus H$. For all sufficiently large $n$, we have $|f_n - f_\infty| < \delta/3$. In particular, $f_n(x_\infty) > f_\infty(x_\infty) - \delta/3$. On the other hand, if $x \in G \setminus H$, then

$$f_n(x) < f_\infty(x) + \delta/3 < f_\infty(x_\infty) - 2\delta/3.$$

Thus, $\arg\max f_n \in H$. Furthermore, $f_n(x_n) \geq f_n(x_\infty) > f_\infty(x_\infty) - \delta/3$. In the other direction, $f_\infty(x_\infty) \geq f_\infty(x_n) > f_n(x_n) - \delta/3$. So

$$|\max f_n - \max f_\infty| < \delta/3,$$

which completes the proof. □

For the final lemma, consider a population consisting of $n$ objects. Suppose $r$ are red, and $r/n \to \rho$ with $0 < \rho < 1$. (The remaining $n - r$ objects are colored black.) Now choose $m$ out of the $n$ objects at random without replacement, where $m/n \to \lambda$ with $0 < \lambda < 1$. Let $X_m$ be the number of red objects that are chosen. So $X_m$ is hypergeometric. The lemma puts no conditions on the joint distribution of the $\{X_m\}$. Only the marginals are relevant.

LEMMA 5. $X_m/n \to \lambda\rho$ almost surely as $n \to \infty$.

PROOF. Of course, $E(X_m) = rm/n$. The lemma can be proved by using Chebychev's inequality, after showing that

$$E\left[\left(X_m - r\frac{m}{n}\right)^4\right] = O(n^2).$$

Tedious algebra can be reduced by appealing to Theorem 4 in Hoeffding (1963). In more detail, let $W_i$ be independent 0–1 variables with $P(W_i = 1) = r/n$. Thus, $\sum_{i=1}^m W_i$ is the number of reds in $m$ draws with replacement, while $X_m$ is the number of reds in $m$ draws without replacement. According to Hoeffding's theorem, $X_m$ is more concentrated around the common expected value. In particular,

$$E\left\{\left(X_m - r\frac{m}{n}\right)^4\right\} < E\left\{\left[\sum_{i=1}^m \left(W_i - \frac{r}{n}\right)\right]^4\right\}.$$

Expanding $[\sum_{i=1}^m (W_i - \frac{r}{n})]^4$ yields $m$ terms of the form $(W_i - \frac{r}{n})^4$. Each of these terms is bounded above by 1. Next consider terms like $(W_i - \frac{r}{n})^2(W_j - \frac{r}{n})^2$ with $i \neq j$. The number of such terms is of order $m^2$, and each term is bounded above by 1. All remaining terms have expectation 0. Thus, $E[(X_n - r\frac{m}{n})^4]$ is of order $m^2 < n^2$. □

NOTE. There are $m^4$ terms in $(a_1 + \cdots + a_m)^4 = \sum_{ijk\ell} a_i a_j a_k a_\ell$. By combinatorial arguments:

(i) $m$ terms are like $a_i^4$, with one index only.

(ii) $3m(m-1)$ are like $a_i^2 a_j^2$, with two different indices.

(iii) $4m(m-1)$ are like $a_i^3 a_j$, with two different indices.

(iv) $6m(m-1)(m-2)$ are like $a_i^2 a_j a_k$, with three different indices.

(v) $m(m-1)(m-2)(m-3)$ are like $a_i a_j a_k a_\ell$, with four different indices.

The counts can also be derived from the "multinomial theorem," which expands $(a_1 + \cdots + a_m)^N$. For an early—and very clear—textbook exposition, see Chrystal (1889, pages 14–15). A little care is needed, since our counts do not restrict the order of the indices: $i < j$ and $i > j$ are both allowed. By contrast, in the usual statements of the multinomial theorem, indices are ordered ($i < j$). German scholarship traces the theorem ("der polynomische Lehrsatz") back to correspondence between Leibniz and Johann Bernoulli in 1695; see, for instance, Tauber (1963), Netto (1927, page 58), and Tropfke (1903, page 332). On the other hand, de Moivre (1697) surely deserves some credit.

We return now to our main objectives. In outline, we must show that $L_n(\beta)/n$ converges to a limit $L_\infty(\beta)$, uniformly over $\beta$ in any bounded set; this will follow from Lemma 5. The limiting $L_\infty(\beta)$ is a strictly concave function of $\beta$, with a unique maximum at $\beta_\infty$: see Proposition 1. Furthermore, $\hat{\beta}_n \to \beta_\infty$ by Lemma 4. In principle, randomization ensures that the balance conditions are nearly satisfied, so the plug-in estimator is consistent by Lemmas 1–2. A rigorous argument gets somewhat intricate; one difficulty is showing that remote $\beta$'s can be ignored, and Lemma 3 helps in this respect.

Some regularity conditions are needed. Technicalities will be minimized if we assume that $Z_i$ takes only a finite number of values; notational overhead is reduced even further if $Z_i = 0, 1,$ or 2. There are now $3 \times 2 \times 2 = 12$ possible values for the triples $Z_i, Y_i^C, Y_i^T$. We say that subject $i$ is of *type zct* provided

$$Z_i = z, \quad Y_i^C = c, \quad Y_i^T = t.$$

Let $\theta_{z,c,t}$ be the fraction of subjects that are of type $zct$; the number of these subjects is $n\theta_{z,c,t}$.

The $\theta$'s are population-level parameters. They are not random. They sum to 1. We assume the $\theta$'s are all positive. Recall that $\pi_T$ is the fraction of subjects assigned to $T$. This is fixed (not random), and $0 < \pi_T < 1$. The fraction assigned to $C$ is $\pi_C = 1 - \pi_T$. In principle, $\pi_T$, $\pi_C$, and the $\theta_{z,c,t}$ depend on $n$. As $n$ increases, we assume these quantities have respective limits $\lambda_T$, $\lambda_C$ and $\lambda_{z,c,t}$, all positive. Since $z$ takes only finitely many values, $\sum_{z,c,t} \lambda_{z,c,t} = 1$.

When $n$ is large, within type $zct$, the fraction of subjects assigned to $T$ is random, but essentially $\lambda_T$: such subjects necessarily have response $Y_i = t$. Likewise, the fraction assigned to $C$ is random, but essentially $\lambda_C$: such subjects necessarily have response $Y_i = c$. In the limit, the $Z$'s are exactly balanced between $T$ and $C$ within each type of subject. That is the essence of the argument; details follow.

Within type $zct$, let $n_{z,c,t}^T$ and $n_{z,c,t}^C$ be the number of subjects assigned to $T$ and $C$, respectively. So

$$n_{z,c,t}^T + n_{z,c,t}^C = n\theta_{z,c,t}.$$

The variables $n_{z,c,t}^T$ are hypergeometric. They are unobservable. This is because type is unobservable: $Y_i^C$ and $Y_i^T$ are not simultaneously observable.

To analyze the log-likelihood function $L_n(\beta)$, recall that $Y_i = X_i Y_i^T + (1 - X_i) Y_i^C$ is the observed response. Let $n_{z,x,y}$ be the number of $i$ with $Z_i = z, X_i = x, Y_i = y$; here $z = 0, 1$, or $2$, $x = 0$ or $1$, and $y = 0$ or $1$. The $n_{z,x,y}$ are observable because $Y_i$ is observable. They are random because $X_i$ is random. Also let $n_{z,x} = n_{z,x,0} + n_{z,x,1}$, which is the number of subjects $i$ with $Z_i = z$ and $X_i = x$. Now $L_n(\beta)/n$ in (12) is the sum

(19a) $$\sum_{z,x} T_{z,x},$$

where

(19b) $$T_{z,x} = -\frac{n_{z,x}}{n} \log[1 + \exp(\beta_1 + \beta_2 x + \beta_3 z)] + \frac{n_{z,x,1}}{n}(\beta_1 + \beta_2 x + \beta_3 z).$$

(Again, $T$ is for "term," not "treatment.") This can be checked by grouping the terms $T_i$ in (12) according to the possible values of $(Z_i, X_i, Y_i)$. There are six terms $T_{z,x}$ in (19), corresponding to $z = 0, 1$, or $2$ and $x = 0$ or $1$.

We claim

(20) $$n_{z,x,y} = n_{z,0,y}^T + n_{z,1,y}^T \quad \text{if } x = 1,$$
$$= n_{z,y,0}^C + n_{z,y,1}^C \quad \text{if } x = 0.$$

The trick is seeing through the notation. For instance, take $x = 1$. By definition, $n_{z,1,y}$ is the number of $i$ with $Z_i = z, X_i = 1, Y_i = y$. The $i$'s with $X_i = 1$ correspond to subjects in the treatment group, so $Y_i = Y_i^T$. Thus, $n_{z,1,y}$ is the number of $i$ with $Z_i = z, X_i = 1, Y_i^T = y$. Also by definition, $n_{z,c,y}^T$ is the number of subjects with $Z_i = z, X_i = 1, Y_i^C = c, Y_i^T = y$. Now add the numbers for $c = 0, 1$: how these subjects would have responded to the control regime is at this point irrelevant. A similar argument works if $x = 0$, completing the discussion of (20).

Recall that $\theta_{z,c,y} \to \lambda_{z,c,y}$ as $n \to \infty$. Let

$$\theta_z = \sum_{c,y} \theta_{z,c,y} \quad \text{and} \quad \lambda_z = \sum_{c,y} \lambda_{z,c,y}.$$

Thus, $\theta_z$ is the fraction of subjects with $Z_i = z$, and $\theta_z \to \lambda_z$ as $n \to \infty$.

As $n \to \infty$, we claim that

(21) $$n_{z,1,y}/n \to \lambda_T(\lambda_{z,0,y} + \lambda_{z,1,y}),$$

(22) $$n_{z,1}/n \to \lambda_T \lambda_z,$$

(23) $$n_{z,0,y}/n \to \lambda_C(\lambda_{z,y,0} + \lambda_{z,y,1}),$$

(24) $$n_{z,0}/n \to \lambda_C \lambda_z,$$

where, for instance, $\lambda_T$ is the limit of $\pi_T$ as $n \to \infty$. More specifically, there a set $\mathcal{N}$ of probability 0, and (21)–(24) hold true outside of $\mathcal{N}$. Indeed, (21) follows from (20) and Lemma 5. Then (22) follows from (21) by addition over $y = 0, 1$. The last two lines are similar to the first two.

A little more detail on (21) may be helpful. What is the connection with Lemma 5? Consider $n_{z,0,y}^T$, which is the number of subjects of type $z0y$ that are assigned to $T$. The "reds" are subjects of type $z0y$, so the fraction of reds in the population converges to $\lambda_{z,0,y}$, by assumption. We are drawing $m$ times at random without replacement from the population to get the treatment group, and $m/n \to \lambda_T$, also by assumption. Now $X_m$ is the number of reds in the sample, that is, the number of subjects of type $z0y$ assigned to treatment. The lemma tells us that $X_m \to \lambda_T \lambda_{z,0,y}$ almost surely. The same argument works for $n_{z,1,y}^T$. Add to get (21).

Next, fix a positive, finite, real number $B$. Consider the open, bounded, convex polyhedron $G_B$ defined by the six inequalities

(25) $$|\beta_1 + \beta_2 x + \beta_3 z| < B$$

for $x = 0, 1$ and $z = 0, 1, 2$. As $n \to \infty$, we claim that $L_n(\beta)/n \to L_\infty(\beta)$ uniformly over $\beta \in G_B$, where

(26a) $L_\infty(\beta) = \lambda_T \Lambda_T + \lambda_C \Lambda_C,$

TABLE 2
*Asymptotic distribution of $\{Z, X, Y\}$ expressed in terms of $\lambda_T, \lambda_C$ and $\lambda_{z,c,t}$*

| Value | Weight |
|---|---|
| $z11$ | $\lambda_T(\lambda_{z,0,1} + \lambda_{z,1,1})$ |
| $z10$ | $\lambda_T(\lambda_{z,0,0} + \lambda_{z,1,0})$ |
| $z01$ | $\lambda_C(\lambda_{z,1,0} + \lambda_{z,1,1})$ |
| $z00$ | $\lambda_C(\lambda_{z,0,0} + \lambda_{z,0,1})$ |

(26b)
$$\Lambda_T = \sum_z (-\lambda_z \log[1 + \exp(\phi_T(z))] + (\lambda_{z,0,1} + \lambda_{z,1,1})\phi_T(z)),$$

(26c)
$$\Lambda_C = \sum_z (-\lambda_z \log[1 + \exp(\phi_C(z))] + (\lambda_{z,1,0} + \lambda_{z,1,1})\phi_C(z)),$$

(26d)
$$\phi_T(z) = \beta_1 + \beta_2 + \beta_3 z,$$
$$\phi_C(z) = \beta_1 + \beta_3 z.$$

(Recall that $\lambda_T$ was the limit of $\pi_T$ as $n \to \infty$, and likewise for $\lambda_C$.) This follows from (21)–(24), on splitting the sum in (19) into two sums, one with terms $z1$ and the other with terms $z0$. The $z1$ terms give us $\lambda_T \Lambda_T$, and the $z0$ terms give us $\lambda_C \Lambda_C$. The conclusion holds outside the null set $\mathcal{N}$ defined for (21)–(24).

It may be useful to express the limiting distribution of $\{Z, X, Y\}$ in terms of $\lambda_T, \lambda_C$ and $\lambda_{z,c,t}$, the latter being the limiting fraction of subjects of type $zct$. See Table 2. For example, what fraction of subjects have $Z = z, X = 1, Y = 1$ in the limit? The answer is the first row, second column of the table. The other entries can be read in a similar way.

The function $\beta \to L_\infty(\beta)$ is strictly concave, by Proposition 1 with $n = 12$ and $p = 3$. The rows of $(My)$ run through all 12 combinations of $1 z x y$ with $z = 1, 2, 3$, and $x = 0, 1$, and $y = 0, 1$. The weights are shown in Table 2.

Let $\beta_\infty$ be the $\beta$ that maximizes $L_\infty(\beta)$. Choose $B$ in (25) so large that $\beta_\infty \in G_B$. Lemma 4 shows that $\max_{\beta \in G_B} L_n(\beta)/n$ is close to $L_\infty(\beta_\infty)$ for all large $n$. Outside $G_B$—if $B$ is large enough—$L_n(\beta)/n$ is too small to matter; additional detail is given below. Thus, $\hat{\beta}_n \in G_B$ for all large $n$, and converges to $\beta_\infty$.

This completes the argument for (14) and we turn to proving (15)—the consistency of the plug-in estimators defined by (10). Recall that $\theta_z$ is the fraction of $i$'s with $Z_i = z$; and $\theta_z \to \lambda_z$ as $n \to \infty$. Now

$$\tilde{\alpha}^T = \frac{1}{n} \sum_{i=1}^n p(\hat{\beta}_n, 1, Z_i)$$
$$= \sum_z \theta_z p(\hat{\beta}_n, 1, z)$$
$$\to \sum_z \lambda_z p(\beta_\infty, 1, z),$$

where the function $p(\beta, x, z)$ was defined in (4). Remember, $z$ takes only finitely many values! A similar argument shows that $\tilde{\alpha}^C \to \sum_z \lambda_z p(\beta_\infty, 0, z)$.

The limiting distribution for $\{Z_i, Y_i^C, Y_i^T\}$ is defined by the $\lambda_{z,c,t}$, where $\lambda_{z,c,t}$ is the limiting fraction of subjects of type $zct$; recall that $\lambda_z = \sum_{c,t} \lambda_{z,c,t}$. We claim

(27) $$\sum_z \lambda_z p(\beta_\infty, 1, z) = \sum_{z,c} \lambda_{z,c,1},$$

(28) $$\sum_z \lambda_z p(\beta_\infty, 0, z) = \sum_{z,t} \lambda_{z,1,t}.$$

Indeed, (22) and (24) show that in the limit, the $Z_i$ are exactly balanced between $T$ and $C$. Likewise, (21) and (23) show that in the limit, the pairs $Y_i^T, Y_i^C$ are exactly balanced between $T$ and $C$. Apply Lemmas 1–2. The left-hand side of (27) is the plug-in estimator for the limiting $\alpha^T$. The right-hand side is the ITT estimator, as well as truth. The three values coincide by the lemmas. The argument for (28) is the same, completing the discussion of (27)–(28).

The right-hand side of (27) can be recognized as the limit of $\frac{1}{n} \sum_{i=1}^n Y_i^T = \sum_{z,c} \theta_{z,c,1}$; likewise, the right-hand side of (28) is the limit of $\frac{1}{n} \sum_{i=1}^n Y_i^C$. This completes the proof of (15). In effect, the argument parlays Fisher consistency into almost-sure consistency, the exceptional null set being the $\mathcal{N}$ where (21)–(24) fail.

Our results give an indirect characterization of $\lim \beta_n$ as the $\beta$ at which the limiting log-likelihood function (26) takes on its maximum. Furthermore, asymptotic normality of $\{n_{z,c,t}^T\}$ entails asymptotic normality of $\hat{\beta}_n$ and the plug-in estimators, but that is a topic for another day.

**Additional Detail on Boundedness**

Consider a $z1$ term in (19). We are going to show that for $B$ large, this term is too small to matter. Fix a small positive $\epsilon$. By (22), for all large $n$,

$$n_{z,1}/n > (1-\epsilon)\lambda_T \lambda_z;$$

by (21),
$$n_{z,1,1}/n < (1+\epsilon)\lambda_T(\lambda_{z,0,1} + \lambda_{z,1,1}).$$
Let $z' = \beta_1 + \beta_2 + \beta_3 z \geq B > 0$. By Lemma 3,
$$\log[1+\exp(z')] > z' + \exp(-z') - \tfrac{1}{2}\exp(-2z') > z'$$
because $z' \geq B > 0$. Our $z1$ term is therefore bounded above for all large $n$ by
$$[-(1-\epsilon)\lambda_z + (1+\epsilon)(\lambda_{z,0,1} + \lambda_{z,1,1})]\,\lambda_T z'.$$
The largeness needed in $n$ depends on $\epsilon$ not $B$.

We can choose $\epsilon > 0$ so small that
$$(1+\epsilon)(\lambda_{z,0,1} + \lambda_{z,1,1}) < (1-2\epsilon)\lambda_z,$$
because $\lambda_{z,0,1} + \lambda_{z,1,1} < \lambda_z$. Our $z1$ term is therefore bounded above by $-\epsilon\lambda_T\lambda_z B$. For $B$ large enough, this term is so negative as to be irrelevant. The argument works because all $\lambda_{z,c,t}$ are assumed positive, and there are only finitely many of them. A similar argument works for $z' = \beta_1 + \beta_2 + \beta_3 z \leq -B$, and for terms $z0$ in (19). These arguments go through outside the null set $\mathcal{N}$ defined for (21)–(24).

### Summing up

It may be useful to summarize the results so far. The parameter $\alpha^T$ is defined in terms of the study population, as the fraction of successes that would be obtained if all members of the population were assigned to treatment; likewise for $\alpha^C$. See (1). The differential log odds $\Delta$ of success is defined by (2). There is a covariate taking a finite number of values. A fraction of the subjects are assigned at random to treatment, and the rest to control. We fit a logit model to data from this randomized controlled experiment, although the model is likely false. The MLE is $\hat{\beta}_n$. ITT and plug-in estimators are defined by (10)–(11).

The size of the population is $n$. This is increasing to infinity. "Types" of subjects are defined by combinations of possible values for the covariate, the response to control, and the response to treatment. We assume that the fraction of subjects assigned to treatment converges to a positive limit, along with the fraction in each type. The parameters $\alpha^T$ and $\alpha^C$ converge too. This may seem a little odd, but $\alpha^T$ and $\alpha^C$ may depend on the study population, hence on $n$.

THEOREM 1. *Under the conditions of this section, if a logit model is fitted to data from a randomized controlled experiment:* (i) *the MLE $\hat{\beta}_n$ converges to a limit $\beta_\infty$;* (ii) *the plug-in estimator $\tilde{\alpha}^T$, the ITT estimator $\hat{\alpha}^T$, and the parameter $\alpha^T$ have a common limit;* (iii) *$\tilde{\alpha}^C$, $\hat{\alpha}^C$, and $\alpha^C$ have a common limit;* (iv) *$\tilde{\Delta}$, $\hat{\Delta}$, and $\Delta$ have a common limit. Convergence of estimators holds almost surely, as the sample size grows.*

### Estimating Individual-Level Parameters

At the beginning of the paper, it was noted that the individual-level parameters $Y_i^T$ and $Y_i^C$ are estimable. The proof is easy. Recall that $X_i = 1$ if $i$ is assigned to treatment, and $X_i = 0$ otherwise; furthermore, $P(X_i = 1) = \pi_T$ is in $(0, 1)$. Then $Y_i X_i / \pi_T$ is an unbiased estimator for $Y_i^T$, and $Y_i(1-X_i)/(1-\pi_T)$ is an unbiased estimator for $Y_i^C$, where $Y_i = X_i Y_i^T + (1-X_i) Y_i^C$ is the observed response.

## 9. AN INEQUALITY

Let subject $i$ have probability of success $p_i$ if treated, $q_i$ if untreated, with $0 < q_i < 1$ and the $q_i$ not all equal. Suppose
$$\frac{p_i}{1-p_i} = \lambda \frac{q_i}{1-q_i}$$
for all $i$, where $\lambda > 1$. Thus,
$$p_i = \frac{\lambda q_i}{1+(\lambda-1)q_i}$$
and $0 < p_i < 1$. Let $\overline{p} = \frac{1}{n}\sum_i p_i$ be the average value of $p_i$, and likewise for $\overline{q}$. We define the *pooled multiplier* as
$$\frac{\overline{p}/(1-\overline{p})}{\overline{q}/(1-\overline{q})}.$$
The log of this quantity is analogous to the differential log odds in (2).

The main object in this section is showing that

(29) $\lambda$ is strictly larger than the pooled multiplier.

Russ Lyons suggested this elegant proof. Fix $\lambda > 1$. Let $f(x) = x/(1-x)$ for $0 < x < 1$. So $f$ is strictly increasing. Let $h(x) = f^{-1}(\lambda f(x))$, so $p_i = h(q_i)$. Inequality (29) says that $f(\overline{p}) < \lambda f(\overline{q})$, that is, $\overline{p} < h(\overline{q})$. Since $p_i = h(q_i)$, proving (29) comes down to proving that $h$ is strictly concave. But
$$h(x) = \frac{\lambda x}{1+(\lambda-1)x}$$
$$= \frac{\lambda}{\lambda-1}\left(1 - \frac{1}{1+(\lambda-1)x}\right),$$
and $y \to 1/y$ is strictly convex for $y > 0$. This completes the proof of (29).

In the other direction,

$$(30) \qquad \frac{\overline{p}}{1-\overline{p}} - \frac{\overline{q}}{1-\overline{q}} = \frac{\overline{p}-\overline{q}}{(1-\overline{p})(1-\overline{q})} > 0$$

because $p_i > q_i$ for all $i$. So the pooled multiplier exceeds 1. In short, given the assumptions of this section, pooling moves the multiplier downward towards 1. Of course, if $\lambda < 1$, we could simply interchange $p$ and $q$. The conclusion: pooling moves the multiplier toward 1.

In this paper, we are interested in estimating differential log odds. If the logit model (4) is right, the coefficient $\beta_2$ of the treatment indicator is a biased estimator of the differential log odds $\Delta$ in (2)—biased away from 0. That is what the inequalities of this section demonstrate, the assumptions being $\beta_3 \neq 0$, $Z_i$ is nonrandom, and $Z_i$ shows variation across $i$. (Random $Z_i$ are easily accommodated.)

If the logit model is wrong, the inequalities show that $\hat{\beta}_2 > \hat{\Delta}$ if $\hat{\Delta} > 0$, while $\hat{\beta}_2 < \hat{\Delta}$ if $\hat{\Delta} < 0$. The assumptions are the same, with $\beta_3$ replaced by $\hat{\beta}_3$, attention being focused on the limiting values defined in the previous section. Since the plug-in estimator $\hat{\Delta}$ is consistent, $\hat{\beta}_2$ must be inconsistent.

The pooling covered by (29)–(30) is a little different from the collapsing discussed in Guo and Geng (1995). (i) Pooling does not involve a joint distribution for $\{X_i, Z_i\}$, or a logit model connecting $Y_i$ to $X_i$ and $Z_i$. (ii) Guo and Geng consider the distribution of one triplet $\{Y_i, X_i, Z_i\}$ only, that is, $n = 1$.

## ACKNOWLEDGMENTS

Thad Dunning, Winston Lim, Russ Lyons, Philip Stark and Peter Westfall made helpful comments, as did an anonymous editor. Ed George deserves special thanks for helpful comments and moral support.

## REFERENCES

Amemiya, T. (1981). Qualitative response models: A survey. *J. Economic Literature* **19** 1483–1536.

Amemiya, T. (1985). *Advanced Econometrics*. Harvard Univ. Press.

Aris, E. M. D., Hagenaars, J. A. P., Croon, M. and Vermunt, J. K. (2000). The use of randomization for logit and logistic models. In *Proceedings of the Fifth International Conference on Social Science Methodology* (J. Blasius, J. Hox, E. de Leuw and P. Smidt, eds.). TT Publications, Cologne.

Berk, R. A. (2004). *Regression Analysis: A Constructive Critique*. Sage, Thousand Oaks, CA.

Berkson, J. (1944). Application of the logistic function to bio-assay. *J. Amer. Statist. Assoc.* **39** 357–365.

Brady, H. E. and Collier, D. (2004). *Rethinking Social Inquiry: Diverse Tools, Shared Standards*. Rowman & Littlefield, Lanham, MD.

Brant, R. (1996). Digesting logistic regression results. *The American Statistician* **50** 117–119.

Chrystal, G. (1889). *Algebra: An Elementary Text Book for the Higher Classes of Secondary Schools and for Colleges*. Part II. Adam and Charles Black, Edinburgh. Available on Google Scholar 7/28/07.

Dabrowska, D. M. and Speed, T. P. (1990). On the application of probability theory to agricultural experiments. Essay on principles. Section 9. *Statist. Sci.* **5** 456–480. MR1092986

de Moivre, A. (1697). A method of raising an infinite multinomial to any given power, or extracting any given root of the same. *Philos. Trans. Roy. Soc. London* **19** 619–625.

Duch, R. M. and Palmer, H. D. (2004). It's not whether you win or lose, but how you play the game. *American Political Science Review* **98** 437–452.

Ducharme, G. R. and Lepage, Y. (1986). Testing collapsibility in contingency tables. *J. Roy. Statist. Soc. Ser. B* **48** 197–205. MR0867997

Evans, W. N. and Schwab, R. M. (1995). Finishing high school and starting college: Do Catholic schools make a difference? *Quarterly J. Economics* **110** 941–974.

Freedman, D. A. (2005). *Statistical Models: Theory and Practice*. Cambridge Univ. Press. MR2175838

Freedman, D. A. (2006a). Statistical models for causation: What inferential leverage do they provide? *Evaluation Review* **30** 691–713.

Freedman, D. A. (2006b). On the so-called "Huber Sandwich Estimator" and "robust standard errors." *Amer. Statist.* **60** 299–302. MR2291297

Freedman, D. A. (2008a). On regression adjustments to experimental data. *Adv. in Appl. Math.* **40** 180–193. MR2388610

Freedman, D. A. (2008b). On regression adjustments in experiments with several treatments. *Ann. Appl. Statist.* **2** 176–196.

Frey, B. S. and Meier, S. (2004). Social comparisons and pro-social behavior: Testing "conditional cooperation" in a field experiment. *American Economic Review* **94** 1717–1722.

Gail, M. H. (1986). Adjusting for covariates that have the same distribution in exposed and unexposed cohorts. In *Modern Statistical Methods in Chronic Disease Epidemiology* (S. H. Moolgavkar and R. L. Prentice, eds.) 3–18. Wiley, New York.

Gail, M. H. (1988). The effect of pooling across strata in perfectly balanced studies. *Biometrics* **44** 151–162.

Gertler, P. (2004). Do conditional cash transfers improve child health? Evidence from PROGRESA's control randomized experiment. *American Economic Review* **94** 336–341.

Gilens, M. (2001). Political ignorance and collective policy preferences. *American Political Science Review* **95** 379–396.

Guo, G. H. and Geng, Z. (1995). Collapsibility of logistic regression coefficients. *J. Roy. Statist. Soc. Ser. B* **57** 263–267. MR1325390

Heckman, J. J. (2000). Causal parameters and policy analysis in economics: A twentieth century retrospective. *Quarterly J. Economics* **115** 45–97.

Hill, A. B. (1961). *Principles of Medical Statistics*, 7th ed. The Lancet, London.

Hodges, J. L. and Lehmann, E. (1964). *Basic Concepts of Probability and Statistics*. Holden-Day, San Francisco. MR0185709

Hoeffding, H. (1963). Probability inequalities for sums of bounded random variables. *J. Amer. Statist. Assoc.* **58** 13–30. MR0144363

Holland, P. W. (1986). Statistics and causal inference (with discussion). *J. Amer. Statist. Assoc.* **8** 945–970. MR0867618

Hu, W.-Y. (2003). Marriage and economic incentives: Evidence from a welfare experiment. *J. Human Resources* **38** 942–963.

Koch, C. G. and Gillings, D. B. (2005). Inference, design-based vs. model-based. In *Encyclopedia of Statistical Sciences* (S. Kotz, C. B. Read, N. Balakrishnan and B. Vidakovic, eds.), 2nd ed. Wiley, Hoboken, NJ.

Lane, P. W. and Nelder, J. A. (1982). Analysis of covariance and standardization as instances of prediction. *Biometrics* **38** 613–621.

Lim, W. (1999). Estimating impacts on binary outcomes under random assignment. Unpublished technical note, MDRC, New York.

Middleton, J. (2007). Even for randomized experiments, logistic regression is not generally consistent. Unpublished technical note, Political Science Dept., Yale Univ.

Netto, E. (1927). *Lehrbuch der Combinatorik*. Teubner, Leipzig.

Neyman, J. (1923). Sur les applications de la théorie des probabilités aux experiences agricoles: Essai des principes. *Roczniki Nauk Rolniczych* **10** 1–51. (In Polish.) English translation by D. M. Dabrowska and T. P. Speed (1990) *Statist. Sci.* **5** 465–480 (with discussion).

Neyman, J., Kolodziejczyk, S. and Iwaszkiewicz, K. (1935). Statistical problems in agricultural experimentation. *J. Roy. Statist. Soc.* **2** Supplement 107–154.

Pate, A. M. and Hamilton, E. E. (1992). Formal and informal deterrents to domestic violence: The Dade county spouse assault experiment. *American Sociological Review* **57** 691–697.

Pratt, J. W. (1981). Concavity of the log likelihood. *J. Amer. Statist. Assoc.* **76** 103–106. MR0608179

Robins, J. M. (1999). Association, causation, and marginal structural models. *Synthese* **121** 151–179. MR1766776

Robinson, L. D. and Jewell, N. P. (1991). Some surprising results about covariate adjustment in logistic regression models. *Internat. Statist. Rev.* **58** 227–240.

Rosenbaum, P. R. (2002). Covariance adjustment in randomized experiments and observational studies (with discussion). *Statist. Sci.* **17** 286–327. MR1962487

Rosenblum, M. and van der Laan, M. J. (2008). Using regression models to analyze randomized trials: Asymptotically valid hypothesis tests despite incorrectly specified models. Available at http://www.bepress.com/ucbbiostat/paper219/.

Scheffé, H. (1956). Alternative models for the analysis of variance. *Ann. Math. Statist.* **27** 251–271. MR0082249

Tauber, S. (1963). On multinomial coefficients. *Amer. Math. Monthly* **70** 1058–1063. MR0160735

Tropfke, J. (1903). *Geschichte der Elementar-mathematik in systematischer Darstellung*. Verlag Von Veit & Comp, Leipzig.

Truett, J., Cornfield, J. and Kannel, W. (1967). A multivariate analysis of the risk of coronary heart disease in Framingham. *J. Chronic Diseases* **20** 511–524.

Verhulst, P. F. (1845). Recherches mathématiques sur la loi d'accroissement de la population. *Nouveaux mémoires de l'Académie Royale des Sciences et Belles-Lettres de Bruxelles* **18** 1–38.

Yule, G. U. (1925). The growth of population and the factors which control it (with discussion). *J. Roy. Statist. Soc.* **88** 1–62.